%% file: text.tex
\documentclass[12pt]{article}
\usepackage{graphicx}
\begin{document}

\noindent
To appear in {\em Local Structure from Diffraction, \\
Fundamental Materials Science Series,} eds. M.F. Thorpe \\
and S.J.L. Billinge, Plenum Press, New York, 1998.
\vspace{8 \baselineskip} \\
{\bf NON-MEAN-FIELD THEORIES OF SHORT RANGE ORDER AND 
\vspace{0.5 \baselineskip} \\
DIFFUSE SCATTERING ANOMALIES IN DISORDERED ALLOYS}
\vspace{3 \baselineskip} \\
\hspace*{1in}Igor Tsatskis\footnote{Former name: I.V. Masanskii}
\vspace{\baselineskip} \\
\hspace*{1in}Department of Earth Sciences \\
\hspace*{1in}University of Cambridge \\
\hspace*{1in}Downing Street \\
\hspace*{1in}Cambridge CB2 3EQ \\
\hspace*{1in}United Kingdom
\vspace{3 \baselineskip} \\
\input{part1.tex}
\input{part2.tex}
\input{part3.tex}

\end{document}

%% file: part1.tex
{\bf INTRODUCTION} \\

Local, or short-range, order in disordered alloys is an important 
and exciting phenomenon which is quantified in electron, X-ray and 
neutron scattering experiments. It is discussed in many excellent 
reviews and books,$^{1-6}$ as well as in the multitude of original 
research papers. 

This relatively short review of the subject does not attempt to 
discuss all aspects of the problem of local correlations in alloys. 
In particular, we will not touch such issues as multiatom (cluster) 
interactions, static displacements and vibrations of alloy atoms, 
partially ordered, multicomponent or amorphous alloys. As a result, 
we will concentrate on the Hamiltonian traditional for the considered 
problem, that of the Ising model on a rigid ideal lattice with pair, 
but otherwise arbitrary (i.e., of any range) interatomic interactions. 

The central object of the paper is the pair correlation function of 
the corresponding dynamical variables of the model, the occupation 
numbers or spin variables, the Fourier transform of which is 
proportional to the intensity of diffuse scattering caused by atomic 
short-range order. The main aim is to show that the expression for 
this quantity has certain internal structure analogous, e.g., to that 
of the averaged Green's function used in the electronic theory of 
disordered alloys. This structure is independent of the approximation 
used for the quantitative description of correlations. As will be seen, 
this structure alone, without further specification of a particular 
theory of short-range order, allows us to see new possibilities in diffuse 
scattering, some of which have recently been observed experimentally. 

The present contribution is organized as follows. First two sections 
are auxiliary and serve mainly for reference purposes; the former 
introduces relation between the phenomenological Hamiltonian of binary 
solid solutions and the Ising model, as well as necessary definitions 
and formulae, while the latter describes briefly standard approaches to 
the theory of electronic structure of disordered alloys. Readers familiar 
with the material contained in these sections can skip it and proceed 
directly to the third section where the key expression~(\ref{70}) for 
the diffuse intensity is introduced. In the fourth section its derivation 
and relation with alloy thermodynamics are discussed. Without any doubt, 
many readers would quickly realize that Eq.~(\ref{70}) is simply 
one of the possible forms of the famous Dyson equation. Such readers  
can then focus on the two last sections. The fifth section reviews, 
from the point of view adopted in this paper, existing theoretical 
approaches -- both traditional and relatively new -- to the problem of 
calculation of the pair correlation function and the diffuse-scattering 
intensity. Finally, in the sixth section based on the recent author's work 
it is shown how the internal structure of Eq.~(\ref{70}) leads to 
understanding existing and predicting new diffuse-scattering anomalies.
\vspace{2 \baselineskip} \\
{\bf LIST OF ABBREVIATIONS} \\

\noindent
\makebox[2cm][l]{AE} alpha-expansion \\
\makebox[2cm][l]{ATA} average $t$-matrix approximation \\
\makebox[2cm][l]{CPA} coherent potential approximation \\
\makebox[2cm][l]{GEM} gamma-expansion method \\
\makebox[2cm][l]{HTE} high-temperature expansion \\
\makebox[2cm][l]{KCM} Krivoglaz-Clapp-Moss (approximation) \\
\makebox[2cm][l]{MC} Monte Carlo (simulation) \\
\makebox[2cm][l]{MFA} mean-field approximation \\
\makebox[2cm][l]{PCF} pair correlation function \\
\makebox[2cm][l]{RPA} random-phase approximation \\
\makebox[2cm][l]{SM} spherical model \\
\makebox[2cm][l]{SRO} short-range order \\
\makebox[2cm][l]{SSA} single-site approximation \\
\makebox[2cm][l]{VCA} virtual crystal approximation 
\vspace{2 \baselineskip} \\
{\bf DESCRIPTION OF A BINARY ALLOY} \\

We will consider the standard model of a binary alloy used 
in the statistical theory of ordering.$^{1-6}$ In this model 
two sorts of atoms (A and B) are distributed over $N$ 
sites of a rigid lattice; there are no vacancies or other 
lattice defects. For simplicity reasons the consideration 
is confined to the case of the lattice with one site per unit 
cell. All sites of the lattice are equivalent, i.e., there 
are no distinct sublattices; this situation corresponds to 
disorder or ferromagnetic ordering. A particular alloy 
configuration is fully described by 
the set of occupation numbers~$p^{\alpha}_{i}$,
\begin{equation}
p^{\alpha}_{i} = \left\{ \begin{array}{ll}
1, & \mbox{atom of type $\alpha$ at the lattice site $i$,} \\
0, & \mbox{otherwise.}
\end{array} \right. \label{1}
\end{equation}
The occupation numbers satisfy obvious relations
\begin{eqnarray}
\sum_{\alpha} p^{\alpha}_{i} & = & 1 \ , \label{2a} \\
\sum_{i} p^{\alpha}_{i} & = & N^{\alpha} \ , \label{2b} \\
\sum_{\alpha} N^{\alpha} & = & N \ , \label{2c}
\end{eqnarray}
where $N^{\alpha}$ is the total number of atoms of type 
$\alpha$, and the last equation holds because the total 
number of alloy atoms is equal to the total 
number of lattice sites. Indices $\alpha$ and $\beta$ take 
only two values, A or B. It is assumed that the interatomic
interactions are pairwise; an atom of type $\alpha$ at site 
$i$ interacts with an atom of type $\beta$ at site $j$ with 
a potential $V^{\alpha \beta}_{ij}$. Then the configurational 
part of the alloy Hamiltonian is
\begin{equation}
H = \frac{1}{2} \sum_{ij} \sum_{\alpha \beta} 
V^{\alpha \beta}_{ij} p^{\alpha}_{i} p^{\beta}_{j} \ . \label{3}
\end{equation}
The main aim of statistical mechanics is the calculation of the 
partition function$^{7}$ 
\begin{equation}
Z = \mbox{Tr} \exp ( - \beta H ) \ , \label{4}
\end{equation}
where $\beta=1/k_{B}T$, $k_{B}$ is the Boltzmann constant, $T$ 
the absolute temperature, and Tr denotes the trace of a matrix. 
It is convenient to avoid restriction~(\ref{2b}) on the total 
number of atoms of each type and work in the grand-canonical 
ensemble, calculating the grand partition function $\tilde{Z}$, 
\begin{equation}
\tilde{Z} = Z \exp \left( \beta \sum_{\alpha}
\mu^{\alpha} N^{\alpha} \right) \ , \label{5}
\end{equation}
where $\mu^{\alpha}$ is the chemical potential of atoms 
$\alpha$. In this approach the number of atoms is a function 
of the corresponding chemical potential, and after having done 
the calculation the chemical potential is adjusted to get the 
required number of atoms. To unify canonical and grand-canonical 
treatments, i.e., to get the same formula~(\ref{4}) for $\tilde{Z}$, 
an effective Hamiltonian $\tilde{H}$ is often introduced:
\begin{equation}
\tilde{H} = H - \sum_{\alpha} \mu^{\alpha} N^{\alpha} \ . \label{6}
\end{equation}
The expression for the effective Hamiltonian is then
\begin{equation}
\tilde{H} = \frac{1}{2} \sum_{ij} \sum_{\alpha \beta}
V^{\alpha \beta}_{ij} p^{\alpha}_{i} p^{\beta}_{j} -
\sum_{i} \sum_{\alpha} \mu^{\alpha} p^{\alpha}_{i} \ . \label{7}
\end{equation}

As is well-known, the statistical-mechanical problem defined 
by the Hamiltonian (\ref{7}) can be transformed into the 
equivalent problem of an Ising magnet in a magnetic field. The 
occupation numbers are not yet all independent, due to 
Eq.~(\ref{2a}). Dependent variables can be eliminated by introducing 
a spin variable $s_{i}$,
\begin{equation}
s_{i} = \left\{ \begin{array}{ll}
+1, & \mbox{spin up (atom $A$) at the site $i$,} \\
-1, & \mbox{spin down (atom $B$) at the site $i$.}
\end{array} \right. \label{8}
\end{equation}
The relations between the occupation numbers and the spin variables are
\begin{eqnarray}
p^{A}_{i} & = & \frac{1}{2} ( 1 + s_{i} ) \ , \label{9a} \\
p^{B}_{i} & = & \frac{1}{2} ( 1 - s_{i} ) \ . \label{9b}
\end{eqnarray}
Inserting Eqs.~(\ref{9a}) and (\ref{9b}) into the expression~(\ref{7}) 
for the effective Hamiltonian $\tilde{H}$, we find that, apart from the
configuration-independent term, the Hamiltonian is that of the 
Ising model,
\begin{equation}
\tilde{H} = - \frac{1}{2} \sum_{ij} J_{ij} s_{i} s_{j} -
h \sum_{i} s_{i} \ , \label{10}
\end{equation}
where the effective exchange integral $J_{ij}$ and the magnetic 
field $h$ are given by
\begin{eqnarray}
J_{ij} & = & - \frac{1}{2} V_{ij} \ , \label{11a} \\
h & = & \frac{1}{2} \left( \mu^{A} - \mu^{B} \right) - \frac{1}{4}
\sum_{j} \left( V^{AA}_{ij} - V^{BB}_{ij} \right) \ , \label{11b} \\
V_{ij} & = & \frac{1}{2} (V^{AA}_{ij}+V^{BB}_{ij}) - 
V^{AB}_{ij} \ , \label{11c}
\end{eqnarray}
and $V_{ij}$ is the pair ordering potential. 
The second term in Eq.~(\ref{11b}) does not, in fact,
depend on the index $i$ because all interatomic potentials are
functions of only the distance between interacting atoms.

For the following we define concentrations $c^{\alpha}$ and 
magnetization $m$ which are statistical averages of the occupation 
numbers~(\ref{1}) and the spin variables~(\ref{8}), respectively,
\begin{eqnarray}
c^{\alpha} & = & \langle p^{\alpha}_{i} \rangle \ , \label{12a} \\
m & = & \langle s_{i} \rangle \ , \label{12b} 
\end{eqnarray}
getting from Eqs.~(\ref{9a}) and (\ref{9b}) 
\begin{equation}
c = \frac{1}{2} ( 1 + m ) \ , \label{13}
\end{equation}
where $c = c^{A} = 1 - c^{B}$. 
Here the symbol $\langle \ldots \rangle$ denotes the 
statistical averaging with the effective Hamiltonian $\tilde{H}$, 
Eq.~(\ref{7}) or (\ref{10}),
\begin{equation}
\langle \ldots \rangle = \frac{ \mbox{Tr} \left[ \ \ldots \ 
\exp \left( - \beta \tilde{H} \right) \right] }{ \mbox{Tr} 
\exp \left( - \beta \tilde{H} \right) } \ . \label{14}
\end{equation}
We also introduce irreducible PCFs for the occupation numbers,
\begin{equation}
G^{\alpha \beta}_{ij} = \langle p^{\alpha}_{i} p^{\beta}_{j} 
\rangle - \langle p^{\alpha}_{i} \rangle \langle p^{\beta}_{j} 
\rangle \ , \label{15}
\end{equation}
and for the spin variables,
\begin{equation}
G_{ij} = \langle s_{i} s_{j} \rangle -
\langle s_{i} \rangle \langle s_{j} \rangle \ . \label{16}
\end{equation}
The notion of irreducibility comes from the diagram technique 
for the Ising model, in the framework of which it can be shown 
that the irreducible PCF does not contain so-called split diagrams 
which are present in the expansion for the correlation function 
$\langle s_{i} s_{j} \rangle$.$^{8}$ Using relations~(\ref{9a}) 
and (\ref{9b}) between the occupation numbers and the spin variables, 
it is easy to obtain the correspondence between the PCFs~(\ref{15}) 
and (\ref{16}),
\begin{equation}
G^{AA}_{ij} = G^{BB}_{ij} = - G^{AB}_{ij} =
\frac{1}{4} G_{ij} \ , \label{17}
\end{equation}
and, therefore, there exists only one independent PCF for the 
occupation numbers. In the theory of alloys the Warren-Cowley 
parameters $\alpha_{ij}$, instead of the PCFs~(\ref{15}), are 
traditionally used for the description of SRO. They are 
defined as
\begin{equation}
\alpha_{ij} = 1 - \frac{\langle p^{A}_{i}
p^{B}_{j} \rangle}{c(1-c)} \ , \label{18}
\end{equation}
and, according to Eqs.~(\ref{12a}), (\ref{13}), (\ref{15}) and (\ref{17}),
\begin{equation}
G_{ij} = 4c(1-c)\alpha_{ij} = (1-m^{2})\alpha_{ij} \ . \label{19}
\end{equation}

The diagonal matrix elements of
the PCFs (\ref{15}) and (\ref{16}) can be expressed in terms
of the averages~(\ref{12a}) and (\ref{12b}). Taking into account 
identities $(p^{\alpha}_{i})^{2}=p^{\alpha}_{i}$ and $s_{i}^{2}=1$
which follow from the definitions~(\ref{1}) and (\ref{8}), we have
\begin{eqnarray}
G^{AA}_{ii} \ = \ G^{BB}_{ii} \! & = & \! - G^{AB}_{ii} \ = \  
c(1-c) \ , \label{21} \\
G_{ii} & = & 1-m^{2} \ , \label{22a} \\
\alpha_{ii} & = & 1 \ . \label{22b}
\end{eqnarray}
The last equation leads to the important sum rule in the 
reciprocal space. Let us introduce the lattice Fourier transformation,
\begin{eqnarray}
f({\bf k}) & = & \frac{1}{N} \sum_{ij} f_{ij} \exp \left( i {\bf k}
{\bf r}_{ij} \right) \ , \label{23a} \\
f_{ij} & = & \frac{1}{\Omega} \int d {\bf k} \ f({\bf k}) \exp 
\left( -i {\bf k} {\bf r}_{ij} \right) \ , \label{23b}
\end{eqnarray}
where $f_{ij}$ is an arbitrary lattice function, ${\bf k}$ the wavevector, 
${\bf r}_{ij}={\bf r}_{i}-{\bf r}_{j}$, ${\bf r}_{i}$ the radius-vector 
of the site $i$, and the integration in Eq.~(\ref{23b}) is carried 
out over the Brillouin zone of volume $\Omega$. 
Eq.~(\ref{22b}) can then be rewritten as
\begin{equation}
\frac{1}{\Omega} \int d {\bf k} \ \alpha({\bf k}) = 1 \ . \label{24}
\end{equation}
Recalling that the Fourier transform $\alpha({\bf k})$ of the SRO 
parameters $\alpha_{ij}$ is the SRO part of the diffuse-scattering 
intensity measured in Laue units,$^{5}$ we conclude that the sum 
rule~(\ref{24}) expresses the property of conservation of the 
integrated SRO intensity.
\vspace{2 \baselineskip} \\
{\bf ELECTRONIC THEORY OF DISORDERED ALLOYS} \\

In what follows we will occasionally turn to ideas which were 
developed in the theory of electronic structure of disordered 
alloys.$^{2,6,9-11}$ With this in mind, we deviate now from the 
main theme of this paper and consider briefly (and rather 
formally, without discussing the physical meaning of derived
formulae) methods of calculating the Green's function of electrons 
averaged over possible alloy configurations. There is some overlapping 
of notations used in this section with those in the rest of the 
paper; however, the present section is quite isolated from the 
statistical-mechanical part of the discussion, and hence
this overlapping should not lead to any confusion.

The central role in the one-electron theory of disordered 
alloys is played by the electronic Green's function $G$ 
which may be defined as a resolvent of the alloy Hamiltonian $H$,
\begin{equation}
G = ( E - H )^{-1} \ , \label{27}
\end{equation}
where $E$ is the energy of an electron. The Hamiltonian of 
a disordered binary alloy is usually assumed to be a sum of 
two operators,
\begin{equation}
H = H_{0} + V \ . \label{28}
\end{equation}
The first of these two terms, $H_{0}$, is translationally 
invariant, while the second, the random one-electron potential $V$, 
depends on particular alloy configuration and is usually 
assumed to be diagonal in the site representation. The operator 
$V$ is a sum of individual potentials $V_{i}$ centred 
at each site $i$ and acquiring two possible values, $V^{A}$ 
and $V^{B}$, in accordance with the type of atom occupying 
the considered site. 
\vspace{\baselineskip} \\
{\bf Propagator expansion} \\

The potential $V$ is often viewed as a perturbation, though not 
necessarily small, of the initial unperturbed Hamiltonian 
$H_{0}$. Defining the unperturbed Green's function,
\begin{equation}
G_{0} = (E-H_{0})^{-1} \ , \label{29}
\end{equation}
one can easily construct the Dyson equation,
\begin{equation}
G = G_{0} + G_{0} V G \ . \label{30}
\end{equation}
Iterating this equation and averaging over all possible alloy
configurations (the averages are denoted by brackets), we get 
the propagator expansion
\begin{equation}
\langle G \rangle = G_{0} + G_{0} \langle V \rangle G_{0} +
G_{0} \langle V G_{0} V \rangle G_{0} + \ldots \ . \label{31}
\end{equation}
The averaged Green's function satisfies another Dyson equation,
\begin{equation}
\langle G \rangle = G_{0} + G_{0} \Sigma
\langle G \rangle \ , \label{32}
\end{equation}
where the operator $\Sigma$ is called the self-energy. The self-energy 
is, in terms of the diagrammatic expansion for the averaged Green's 
function $\langle G \rangle$ generated by Eq.~(\ref{31}), the 
irreducible part of $\langle G \rangle$, i.e., the sum of all 
graphs for $\langle G \rangle$ which cannot be separated into 
two parts by cutting a single bare-propagator line $G_{0}$. From
Eq.~(\ref{32}) it follows that
\begin{equation}
\langle G \rangle = ( G_{0}^{-1}  - \Sigma )^{-1} =
( E - H_{0}  - \Sigma )^{-1} \ , \label{33}
\end{equation}
the second equation comes from Eq.~(\ref{29}).
\vspace{\baselineskip} \\
{\bf Locator expansion} \\

There exists another perturbation series, the locator expansion, 
which is the expansion in powers of the unperturbed Hamiltonian 
$H_{0}$ rather than the potential $V$. In this case the unperturbed 
Green's function is that of the localized atomic states,
\begin{equation}
g = (E-V)^{-1} \ , \label{34}
\end{equation}
and the corresponding Dyson equation has the form
\begin{equation}
G = g + g H_{0} G \ . \label{35}
\end{equation}
As in the case of the propagator expansion, we iterate this 
equation and average over the ensemble of configurations term 
by term, obtaining the following expansion:
\begin{equation}
\langle G \rangle = \langle g \rangle + \langle g H_{0} g \rangle +
\langle g H_{0} g H_{0} g \rangle + \ldots \ . \label{36}
\end{equation}
By introducing the so-called fully renormalized interactor $U$,
\begin{equation}
U = H_{0} + H_{0} G H_{0} \ , \label{37}
\end{equation}
full formal analogy between the propagator 
(Eq.~(\ref{31})) and locator expansions is achieved:
\begin{equation}
\langle U \rangle = H_{0} + H_{0} \langle g \rangle H_{0} +
H_{0} \langle g H_{0} g \rangle H_{0} + \ldots \ . \label{38}
\end{equation}
The corresponding sum of all irreducible graphs $\sigma$ in 
the case of the locator expansion is called the locator; 
similarly to Eq.~(\ref{32}), we have
\begin{equation}
\langle U \rangle = H_{0} + 
H_{0} \sigma \langle U \rangle \ . \label{39}
\end{equation}
From Eqs.~(\ref{37}) and (\ref{39}) it immediately follows 
that in terms of the locator the Dyson equation and the 
expression for $\langle G \rangle$ are
\begin{eqnarray}
\langle G \rangle & = & \sigma + \sigma H_{0}
\langle G \rangle \ , \label{40} \\
\langle G \rangle & = & 
( \sigma^{-1}  - H_{0} )^{-1} \ , \label{41}
\end{eqnarray}
respectively. The relationship between the locator $\sigma$ and 
the self-energy $\Sigma$, according to Eqs.~(\ref{33}) 
and (\ref{41}), is
\begin{equation}
\sigma = ( E - \Sigma )^{-1} \ . \label{42}
\end{equation}
\vspace{0.01 \baselineskip} \\
{\bf Renormalization of one-electron potential} \\

We now renormalize the potential $V$ subtracting a 
configuration-independent, site-diagonal operator $S$ 
and adding it to the unperturbed Hamiltonian $H_{0}$:
\begin{equation}
H = H_{0} + V = ( H_{0} + S ) + ( V - S ) =
\tilde{H} + \tilde{V} \ . \label{43}
\end{equation}
If one defines the unperturbed Green's function with respect 
to the renormalized Hamiltonian $\tilde{H}$,
\begin{equation}
\tilde{G} = ( E - \tilde{H} )^{-1} \ , \label{44}
\end{equation}
then the Green's functions $G$ and $\tilde{G}$ are related 
by the Dyson equation analogous to Eq.~(\ref{30}),
\begin{equation}
G = \tilde{G} + \tilde{G} \tilde{V} G \ . \label{45}
\end{equation}
As follows from Eqs.~(\ref{33}) and (\ref{43}),
\begin{equation}
\langle G \rangle = ( E - \tilde{H}  -
\tilde{\Sigma} )^{-1} \ , \label{46}
\end{equation}
where $\tilde{\Sigma}$ relates to $\tilde{H}$ and $\tilde{V}$ 
in the same way as $\Sigma$ to $H_{0}$ and $V$, and
\begin{equation}
\Sigma = S + \tilde{\Sigma} \ . \label{47}
\end{equation}
The operator $S$ may be regarded as an initial approximation 
for the exact self-energy~$\Sigma$.

The next step is to introduce the total scattering operator $T$ 
and express the exact Green's function $G$ and the self-energy 
$\Sigma$ in terms of $T$. The total scattering operator is 
defined by the relation
\begin{equation}
T = \tilde{V} + \tilde{V} \tilde{G} T \ , \label{48}
\end{equation}
which gives
\begin{equation}
T = ( \tilde{V}^{-1}  - \tilde{G} )^{-1} \ . \label{49}
\end{equation}
Excluding $\tilde{V}$ from Eqs.~(\ref{45}) and (\ref{49}) 
and averaging the result over realizations of the random 
potential, we get the relation between $\langle G \rangle$ and 
$\langle T \rangle$:
\begin{equation}
\langle G \rangle = \tilde{G} + \tilde{G}
\langle T \rangle \tilde{G} \ . \label{50}
\end{equation}
Taking into account the relation~(\ref{33}) between the 
averaged Green's function and the self-energy and using 
Eq.~(\ref{50}), we finally obtain
\begin{equation}
\Sigma = S + ( \langle T \rangle^{-1} + \tilde{G} )^{-1} \ . \label{51}
\end{equation}
Thus, having calculated the averaged total scattering operator
$\langle T \rangle$ one can determine the averaged Green's 
function $\langle G \rangle$ and the self-energy $\Sigma$ 
according to Eqs.~(\ref{50}) and (\ref{51}). The problem 
of the description of the disordered alloy is therefore reduced 
to the problem of finding reasonable approximate expression for 
the operator $\langle T \rangle$. The operator $T$ can be 
expressed in terms of operators describing scattering on 
individual atomic potentials $V_{i}$. To do this, it is 
convenient to decompose the Green's function $\tilde{G}$ into 
two parts which are diagonal and off-diagonal in the site 
representation, respectively:
\begin{equation}
\tilde{G} = \tilde{G}_{d} + \tilde{G}_{od} \ . \label{52}
\end{equation}
Inserting this equation into Eq.~(\ref{49}), we have after 
some straightforward algebra,
\begin{equation}
T = ( t^{-1} - \tilde{G}_{od} )^{-1} \ , \label{53}
\end{equation}
where the operator
\begin{equation}
t = ( \tilde{V}^{-1} - \tilde{G}_{d} )^{-1} \label{54}
\end{equation}
is, similarly to the potential $V$, site-diagonal and 
represents a sum of individual scattering operators $t_{i}$ 
corresponding to atomic potentials.
\vspace{\baselineskip} \\
{\bf Single-site approximations} \\

At any level of approximation there exist two main approaches 
to the problem of calculation of the averaged total scattering 
operator $\langle T \rangle$ and, consequently, the averaged 
Green's function $\langle G \rangle$ and the self-energy $\Sigma$. 
First, it is possible to choose the operator $S$ from the very 
beginning, and then calculate these three quantities using 
Eqs.~(\ref{49})-(\ref{51}); this is the non-self-consistent 
scheme. In most cases the VCA choice 
$S=\langle V \rangle$ is used. Second, one can consider $S$ as 
an operator variable and, noticing that the scattering operator 
$T$ is a function of $S$, determine the latter as a solution
of the equation
\begin{equation}
\langle T[S] \rangle = 0 \ . \label{55}
\end{equation}
This equation gives, according to Eqs.~(\ref{50}) and (\ref{51}),
\begin{eqnarray}
\langle G \rangle & = & \tilde{G} \ , \label{56} \\ 
\Sigma & = & S \ , \label{57}
\end{eqnarray}
and such an approach is called self-consistent. To make the 
calculation of the averaged total scattering operator practically 
possible, the following decoupling of the configurationally-averaged 
Eq.~(\ref{53}), called the SSA, is usually adopted:
\begin{equation}
\langle T \rangle = \langle ( t^{-1} - \tilde{G}_{od} )^{-1} \rangle
\stackrel{SSA}{\longrightarrow} ( \langle t \rangle^{-1} - \tilde{G}_{od}
)^{-1} \ . \label{58}
\end{equation}
In the framework of the SSA Eq.~(\ref{51}) for the 
self-energy takes the form
\begin{equation}
\Sigma^{SSA} = S + ( \langle t \rangle^{-1} +
\tilde{G}_{d} )^{-1} \ , \label{59}
\end{equation}
which means that in this approximation the self-energy is diagonal
in the site representation. Consider now the two approaches 
in combination with the SSA. The non-self-consistent one, as is 
seen from Eqs.~(\ref{54}) and (\ref{59}), gives the self-energy as
a sum of the operator $S$ and the effective scattering potential
$\tilde{V}_{eff}$ corresponding to the average scattering operator
$\langle t \rangle$,
\begin{equation}
\tilde{V}_{eff} = ( \langle t \rangle^{-1} +
\tilde{G}_{d} )^{-1} \ ; \label{60}
\end{equation}
accordingly, this method of calculation is known as the ATA. 
The application of the SSA to the self-consistent scheme 
(Eq.~(\ref{55})) gives the following equation for the
evaluation of the self-energy:
\begin{equation}
\langle t[S] \rangle = 0 \ . \label{61}
\end{equation}
This equation determines the CPA.
\vspace{2 \baselineskip} \\

%% file: part2.tex
{\bf EXPRESSION FOR SHORT-RANGE ORDER DIFFUSE INTENSITY} \\

We now derive a formally exact expression for the SRO part 
$\alpha({\bf k})$ of the diffuse-scattering intensity. 
A key observation here is that the spin PCF~(\ref{16}) satisfies 
the Dyson equation,$^{8}$
\begin{equation}
G = \sigma + \sigma \Delta G \ , \label{62}
\end{equation}
where $\Delta = \beta J$ and, like in the previous section, matrix 
notations are used. Here $\sigma$ is again the sum of all irreducible 
graphs in the diagrammatic expansion for $G$, but irreducibility is 
now defined with respect to the interaction-to-temperature ratio $\Delta$.
Eq.~(\ref{62}) has the same form as the Dyson equation~(\ref{40}) for 
the Green's function of electrons~(\ref{27}) averaged over alloy 
configurations. From Eq.~(\ref{62}) it follows that
\begin{equation}
G = \left( \sigma^{-1} - \Delta \right)^{-1} \ . \label{63}
\end{equation}
This expression for the spin PCF is analogous to Eq.~(\ref{41}).

The irreducible part $\sigma$ of the PCF $G$ is sometimes 
called the self-energy.$^{8}$ However, to maintain the 
analogy with the electronic theory of alloys (i.e., with the 
terminology of the previous section) we will refer to this 
quantity as the locator, and reserve this term for another 
object, defining the PCF self-energy $\Sigma$ by the relation 
similar to Eq.~(\ref{42}): 
\begin{equation}
\sigma = - \Sigma^{-1} \ . \label{64}
\end{equation}
In terms of $\Sigma$ Eq.~(\ref{63}) becomes
\begin{equation}
G = \left( - \Sigma - \Delta \right)^{-1} \ . \label{65}
\end{equation}
We also introduce the locator $\tilde{\sigma}$ and the self-energy
$\tilde{\Sigma}$ for the occupation-number PCFs $G^{\alpha \beta}$
(Eq.~(\ref{15})):
\begin{eqnarray}
\tilde{\sigma} & = & \frac{1}{4} \sigma \ , \label{66} \\
\tilde{\Sigma} & = & 4 \Sigma \ . \label{67}
\end{eqnarray}
Then matrices $\tilde{\sigma}$ and $\tilde{\Sigma}$ are related by 
the same Eq.~(\ref{64}), 
\begin{equation}
\tilde{\sigma} = - \tilde{\Sigma}^{-1} \ , \label{68}
\end{equation}
and we have
\begin{equation}
G^{AA} = G^{BB} = - G^{AB} = c(1-c) \alpha = 
( - \tilde{\Sigma} + 2 \beta V )^{-1} \ . \label{69}
\end{equation}
Written in {\bf k}-representation, this equation leads to the 
following expression for the diffuse-scattering intensity:$^{12,13}$
\begin{equation}
\alpha({\bf k}) = \frac{1}{ c(1-c) \left[ - \tilde{\Sigma}({\bf k}) + 
2 \beta V({\bf k}) \right] } \ . \label{70}
\end{equation}
The central quantity of interest for us here is the PCF self-energy 
$\tilde{\Sigma}({\bf k})$; apart from the wavevector, it depends 
also on two other variables, temperature and concentration.
Later on we will be focusing on existing approximations 
for the PCF self-energy.
\vspace{2 \baselineskip} \\
{\bf SELF-ENERGY AND THERMODYNAMICS} \\

In the last section the Dyson equation~(\ref{62}) and the related 
expression~(\ref{65}) for the spin PCF were simply postulated. However, 
it would be useful to know how Eq.~(\ref{65}) could be derived and how 
the self-energy is related to the thermodynamics of the system.
\vspace{\baselineskip} \\
{\bf Variational formulation of statistical mechanics} \\

As was mentioned before, the main task of the statistical-mechanical 
treatment is to calculate the partition function~(\ref{4}) of a system 
described by the Hamiltonian $H$. The Hamiltonian usually is a linear 
combination,
\begin{equation}
H = \sum_{n} x_{n} a_{n} \ , \label{86}
\end{equation}
of some operators $a_{n}$ with coefficients $x_{n}$. Variables 
$\alpha_{n}$ conjugated to the parameters $x_{n}$ are defined as 
averages~(\ref{14}) of the operators $a_{n}$,
\begin{equation}
\alpha_{n} = \left\langle a_{n} \right\rangle = 
\frac{\partial F}{\partial x_{n}} \ , \label{87}
\end{equation} 
where 
\begin{equation}
F = - k_{B} T \ln Z \label{111}
\end{equation} 
is the free energy of the system, and the second equation in~(\ref{87}) 
comes from the definition~(\ref{4}) and Eq.~(\ref{86}). Our aim now 
is the calculation of the averages $\alpha_{n}$ and the free energy 
$F$ as functions of the parameters $x_{n}$. This problem can be 
formulated as a variational one, if the Legendre transform 
$\Gamma(\alpha)$ of the free energy $F(x)$ is introduced:$^{14,15}$
\begin{equation}
\Gamma(\alpha) = F(x(\alpha)) - 
\sum_{n} \alpha_{n} \, x_{n}(\alpha) \ . \label{88}
\end{equation}
Here the averages $\alpha_{n}$ are the independent variables, and 
$x_{n}(\alpha)$ are solutions of Eqs.~(\ref{87}). Differentiating 
$\Gamma(\alpha)$ with respect to $\alpha_{n}$ and using Eqs.~(\ref{87}), 
we get
\begin{equation}
\frac{\partial \Gamma}{\partial \alpha_{n}} = 
\sum_{m} \frac{\partial F}{\partial x_{m}} \, \frac{\partial x_{m}}{\partial \alpha_{n}} - 
\sum_{m} \left( \frac{\partial \alpha_{m}}{\partial \alpha_{n}} \, x_{m} + 
\alpha_{m} \, \frac{\partial x_{m}}{\partial \alpha_{n}} \right) = - x_{n} \ . \label{89}
\end{equation}
Finally, introducing another function $\Phi(x,\alpha)$, 
\begin{equation}
\Phi(x,\alpha) = \Gamma(\alpha) + \sum_{n} \alpha_{n} x_{n} \ , \label{90}
\end{equation}
which depends on both $\alpha_{n}$ and $x_{n}$, we find from 
Eq.~(\ref{89}) that it is stationary with respect to variations of 
$\alpha_{n}$ at fixed $x_{n}$:
\begin{equation}
\frac{\partial \Phi}{\partial \alpha_{n}} = 0 \ . \label{91}
\end{equation}
At the stationary point $\alpha_{n}=\alpha_{n}(x)$, where $\alpha_{n}(x)$ 
are solutions of Eqs.~(\ref{89}), $\Phi(x,\alpha)$ coincides with the 
free energy $F(x)$:
\begin{equation}
\Phi(x,\alpha(x)) = F(x) \ . \label{92}
\end{equation}
Function $\Phi(x,\alpha)$ is usually called the variational free energy. 
Noting that the internal energy is the statistical average~(\ref{14}) of the 
Hamiltonian, $E = \left\langle H \right\rangle$, $\Phi(x,\alpha)$ can be 
written in the standard thermodynamic form:
\begin{eqnarray}
\Phi(x,\alpha) & = & E(x,\alpha) - T S(\alpha) \ , \label{114a} \\
E(x,\alpha) & = &  \sum_{n} x_{n} \alpha_{n} \ , \label{114b} \\
S(\alpha) & = & - \beta \Gamma(\alpha) \ , \label{114c}
\end{eqnarray}
where $E(x,\alpha)$ and $S(\alpha)$ are the variational internal energy 
and configurational entropy, respectively. Eq.~(\ref{91}) now becomes 
\begin{equation}
T \, \frac{\partial S}{\partial \alpha_{n}} = x_{n} \ . \label{115}
\end{equation}
From Eqs.~(\ref{87}) and (\ref{89}) it follows that  
\begin{equation}
\sum_{l} \frac{\partial^{2} \Gamma}{\partial \alpha_{n} \partial \alpha_{l}} \,
\frac{\partial^{2} F}{\partial x_{l} \partial x_{m}} = 
- \sum_{l} \frac{\partial x_{n}}{\partial \alpha_{l}} \,
\frac{\partial \alpha_{l}}{\partial x_{m}} = 
- \frac{\partial x_{n}}{\partial x_{m}} = - \delta_{nm}  \ . \label{93}
\end{equation}
Eq.~(\ref{93}) shows that matrices of second derivatives of the free 
energy $F(x)$ and its Legendre transform $\Gamma(\alpha)$ (or the 
variational free energy $\Phi(x,\alpha)$ which differs from $\Gamma(\alpha)$ 
only by the bilinear term $\sum_{n} x_{n} \alpha_{n}$) are mutually 
inverse up to a sign.
\vspace{\baselineskip} \\
{\bf First Legendre transformation for the Ising model} \\

We will now apply the general technique of the Legendre transformations 
outlined above to the particular case of the Ising model in the 
inhomogeneous magnetic field. The Hamiltonian of the model is a 
straightforward generalization of Eq.~(\ref{10}):
\begin{equation}
H = - \frac{1}{2} \sum_{ij} J_{ij} s_{i} s_{j} -
\sum_{i} h_{i} s_{i} \ . \label{94}
\end{equation}
From comparison of Eqs.~(\ref{86}) and (\ref{94}) it follows that 
the latter contains two kinds of operators $a_{n}$ - single spin 
variables $s_{i}$ and products $s_{i} s_{j}$ of two spin variables. 
Corresponding parameters $x_{n}$, except for sign, are $h_{i}$ and 
$J_{ij}$, and variables $\alpha_{n}$ conjugated to these parameters 
are $\langle s_{i} \rangle = m_{i}$ (Eq.~(\ref{12b})) and 
$\langle s_{i} s_{j} \rangle = m_{i} m_{j} + G_{ij}$ (Eq.~(\ref{16})). 
In the general case considered before the Legendre transformation 
was carried out with respect to all parameters $x_{n}$; as a result, 
all conjugated variables $\alpha_{n}$ were calculated using the 
variational principle. Here, however, this approach will be applied 
only to the magnetic field $h_{i}$, and only the magnetization $m_{i}$ 
will be calculated by means of the variational procedure. The 
resulting partial transformation is called the first Legendre 
transformation.$^{15-17}$ In this case $h_{i}$ plays the role of 
$x_{n}$, and comparison with Eq.~(\ref{86}) shows that $a_{n}$ 
corresponds to $-s_{i}$. Eqs.~(\ref{87}), (\ref{88})-(\ref{92}) and 
(\ref{93}) now become
\begin{eqnarray}
\Gamma(m,J) & = & F(h(m),J) + \sum_{i} m_{i} \, h_{i}(m) \ , \label{95} \\
\Phi(h,m;J) & = & \Gamma(m,J) -  \sum_{i} m_{i} h_{i} \ , \label{96} \\
\frac{\partial F}{\partial h_{i}} & = & - m_{i} \ , \label{97} \\
\frac{\partial \Gamma}{\partial m_{i}} & = & h_{i} \ , \label{98} \\
\frac{\partial \Phi}{\partial m_{i}} & = & 0 \ , \label{99} \\
\Phi(h,m(h);J) & = & F(h,J) \ , \label{100} \\
\sum_{k} \frac{\partial^{2} \Gamma}{\partial m_{i} \partial m_{k}} \, 
\frac{\partial^{2} F}{\partial h_{k} \partial h_{j}} & = &
\sum_{k} \frac{\partial^{2} \Phi}{\partial m_{i} \partial m_{k}} \, 
\frac{\partial^{2} F}{\partial h_{k} \partial h_{j}} = - \delta_{ij} \ , 
\label{101}
\end{eqnarray}
where, in the same manner as earlier, $h_{i}(m)$ and $m_{i}(h)$ are 
solutions of Eqs.~(\ref{97}) and (\ref{98}), respectively. 
\vspace{\baselineskip} \\
{\bf Derivation of the Dyson equation and meaning of the self-energy} \\

First of all, we note that from the definitions of the partition 
function, the statistical average, the PCF and the free energy 
(Eqs.~(\ref{4}), (\ref{14}), (\ref{16}) and (\ref{111})) it follows 
that for the Hamiltonian~(\ref{94})
\begin{equation}
\frac{\partial^{2} F}{\partial h_{i} \partial h_{j}} = 
- \beta G_{ij} \ . \label{102}
\end{equation}
Then, combining this equation with Eq.~(\ref{101}), we obtain
\begin{equation}
G_{ij} = k_{B} T \left( \frac{\partial^{2} \Phi}{\partial m \partial m} 
\right)^{-1}_{ij} \ . \label{103}
\end{equation}
This notation means that the real-space matrix element of the PCF 
is proportional to the corresponding matrix element of the inverse 
of the matrix whose matrix elements are second derivatives 
$\partial^{2} \Phi / \partial m_{i} \partial m_{j}$ of the variational 
free energy with respect to the corresponding magnetizations. 

The variational free energy can always be written as a sum of its 
mean-field (Bragg-Williams) part and the non-mean-field correction:
\begin{eqnarray}
\Phi & = & \Phi^{MFA} + \delta \Phi \ , \label{104} \\
\Phi^{MFA} & = & E^{MFA} - T S^{MFA} \ . \label{105} 
\end{eqnarray}
The expressions for the mean-field internal energy $E^{MFA}$ and the 
configurational entropy $S^{MFA}$ are well-known:$^{1-6}$
\begin{eqnarray}
E^{MFA} & = & -\frac{1}{2} \sum_{ij} J_{ij} m_{i} m_{j} -
\sum_{i} h_{i} m_{i}  \ , \label{106} \\
S^{MFA} & = & - k_{B} \sum_{i} \left( \frac{1+m_{i}}{2} \ln \frac{1+m_{i}}{2} + 
\frac{1-m_{i}}{2} \ln \frac{1-m_{i}}{2} \right) \ . \label{107} 
\end{eqnarray}
Substituting Eqs.~(\ref{104})-(\ref{107}) into Eq.~(\ref{103}) 
and noticing that 
\begin{equation}
\frac{\partial^{2} E^{MFA}}{\partial m_{i} \partial m_{j}} = 
- J_{ij} \ , \label{108} 
\end{equation}
we recover the result~(\ref{65}) for the PCF (and, therefore, the Dyson 
equation~(\ref{62})) with the following expression for the self-energy:
\begin{equation}
\Sigma_{ij} = \frac{\partial^{2} S^{MFA}}{\partial m_{i} 
\partial m_{j}} - \beta \, \frac{\partial^{2} \left( \delta \Phi 
\right)}{\partial m_{i} \partial m_{j}} \ . \label{109} 
\end{equation}
It is seen that the self-energy is the sum of the second derivatives, 
with respect to the magnetizations, of the two terms contributing to 
the expression for the variational free energy $\Phi$: the mean-field 
configurational entropy and the non-mean-field part of $\Phi$. Noting 
further that 
\begin{equation}
\frac{\partial^{2} S^{MFA}}{\partial m_{i} \partial m_{j}} 
= - \frac{\delta_{ij}}{1-m_{i}^{2}} \ , \label{110} 
\end{equation}
we finally obtain 
\begin{equation}
\Sigma_{ij} = - \frac{\delta_{ij}}{1-m_{i}^{2}} - \beta \, \frac{\partial^{2} 
\left( \delta \Phi \right)}{\partial m_{i} \partial m_{j}} \ . \label{112} 
\end{equation}
The first term in this expression is diagonal in the direct space. 
This means that, back to the homogeneous case in which all the lattice 
sites are equivalent, this term is {\bf k}-independent in the reciprocal 
space, and all the wavevector dependence of the self-energy comes from 
the second term, i.e., is the result of the non-mean-field corrections 
to the MFA variational free energy. 
\vspace{2 \baselineskip} \\
{\bf APPROXIMATIONS FOR THE SELF-ENERGY} \\

In this section our attention will be focused on the formally exact 
result~(\ref{70}) for the SRO diffuse intensity 
$\alpha({\bf k})$, which is the Fourier transform of the Warren-Cowley 
SRO parameters $\alpha_{ij}$~(\ref{18}). Available theories of SRO 
will now be considered, in the light of the structure of Eq.~(\ref{70}), 
as different approximations for the self-energy $\tilde{\Sigma}$.
\vspace{\baselineskip} \\
{\bf Random-phase (Krivoglaz-Clapp-Moss) approximation} \\

The simplest and by far the most popular theory of SRO is the RPA 
(or, in the alloy language, the KCM approximation):$^{5,18,19}$
\begin{equation}
\alpha^{RPA}({\bf k}) = \frac{1}{1 + 2 c(1-c) \beta V({\bf k})} \ . \label{72}
\end{equation}
Eq.~(\ref{72}) is usually derived using mean-field-like arguments.
Comparing Eqs.~(\ref{70}) and (\ref{72}), we see that
\begin{equation}
\tilde{\Sigma}^{RPA}({\bf k}) = - \frac{1}{c(1-c)} \ . \label{73}
\end{equation}
The RPA self-energy~(\ref{73}) is thus wavevector- and 
temperature-independent; it depends only on alloy composition. 
Returning via Eqs.~(\ref{13}) and (\ref{67}) to the magnetic language used 
in the previous section, we conclude that the RPA for the self-energy 
corresponds precisely to neglecting the second term in the right-hand side 
of Eq.~(\ref{112}). The non-mean-field contribution to the self-energy is 
therefore ignored in the RPA, and the self-energy in the direct space is 
simply the second derivative of the mean-field configurational entropy with 
respect to the magnetization:
\begin{equation}
\Sigma^{RPA}_{ij} = \frac{\partial^{2} S^{MFA}}{\partial m_{i} \partial m_{j}} = 
- \frac{\delta_{ij}}{1-m_{i}^{2}} \ . \label{113} 
\end{equation}
From the point of view of the terminology used in the electronic theory of 
disordered alloys, the RPA for the self-energy could be referred to as the 
SSA; indeed, both the RPA (Eq.~(\ref{109})) and the SSA (Eq.~(\ref{59})) 
self-energies are diagonal in the site representation. The RPA resembles 
the non-self-consistent SSA, i.e., the ATA, in the sense that both 
approximations define the corresponding self-energies explicitly. 

The RPA reduces to the well-known Ornstein-Zernike description of 
correlations,$^{7}$ when only those wavevectors which are close to the 
position ${\bf k}_{0}$ of the absolute minimum of the interaction 
$V({\bf k})$ are considered (this approximation corresponds to the 
long-wave limit in the case of ferromagnetic ordering). Let us expand 
$V({\bf k})$ in powers of ${\bf q}={\bf k}-{\bf k}_{0}$ and retain only 
the lowest-order (quadratic) term, 
\begin{equation}
V({\bf k}) \approx V({\bf k}_{0}) + 
\frac{1}{2} \sum_{ij} g_{ij} q_{i} q_{j} \ , \label{74}
\end{equation}
where $g$ is the $3 \times 3$-matrix of second derivatives of 
$V({\bf k})$ at ${\bf k}_{0}$. We will take only the simplest example 
of cubic symmetry, when $g_{ij} = g \delta_{ij}$; in this case 
$\sum_{ij} g_{ij} q_{i} q_{j} = g q^{2}$, where $q \equiv |{\bf q}|$.
Substituting the result into Eq.~(\ref{72}), we obtain
\begin{equation}
\alpha^{RPA}({\bf k}) = \frac{k_{B} T}{k_{B} (T-T_{c}) + 
c(1-c) g q^{2}} \ , \label{75}
\end{equation}
where it is taken into account that the mean-field result for the 
instability temperature $T_{c}$ is 
\begin{equation}
T_{c} = 2 c(1-c) |V({\bf k}_{0})| / k_{B} \ ; \label{76}
\end{equation}
at the position of the absolute minimum the interaction value is negative, 
since the average of the interaction over the Brillouin zone is zero 
(see Eq.~(\ref{80}) below). In real space we get asymptotically
(i.e., at large distances)
\begin{equation}
\alpha^{RPA}_{ij} \propto \frac{1}{r_{ij}} 
\exp (- i {\bf k}_{0} {\bf r}_{ij} - r_{ij} / \xi) \ , \label{77} 
\end{equation}
where $r_{ij} \equiv |{\bf r}_{ij}|$, and  the correlation length 
$\xi$ is
\begin{equation}
\xi = \sqrt{\frac{c(1-c) g}{k_{B}(T-T_{c})}} \ . \label{78}
\end{equation}
Eqs.~(\ref{75})-(\ref{78}) represent the Ornstein-Zernike result 
for the PCF.

However, the RPA expression~(\ref{72}) has a serious disadvantage: it is
unable to satisfy the sum rule~(\ref{24}). Using the identity
\begin{equation}
\frac{1}{1 + 2 c(1-c) \beta V({\bf k})} = 1 - 2 c(1-c) \beta V({\bf k}) + 
\frac{4 c^{2} (1-c)^{2} \beta^{2} V^{2}({\bf k})}{1 + 2 c(1-c) 
\beta V({\bf k})} \ , \label{79}
\end{equation}
it can be shown that the RPA formula~(\ref{72}) always leads to the
overestimation of the value of the integral over the Brillouin zone
(\ref{24}). Since
\begin{equation}
\frac{1}{\Omega} \int d {\bf k} \ V({\bf k}) = V_{ii} = 0 \ , \label{80}
\end{equation}
from Eq.~(\ref{79}) it follows that
\begin{equation}
\frac{1}{\Omega} \int d {\bf k} \ \alpha^{RPA}({\bf k}) = 1 + 
\frac{1}{\Omega} \int d {\bf k} \ \frac{4 c^{2} (1-c)^{2} \beta^{2} 
V^{2}({\bf k})}{1 + 2 c(1-c) \beta V({\bf k})} \ , \label{81}
\end{equation}
and, therefore,
\begin{equation}
\frac{1}{\Omega} \int d {\bf k} \ \alpha^{RPA}({\bf k}) > 1 \ . \label{82}
\end{equation}
This integral is close to unity only at sufficiently high temperatures.
As temperature decreases, the deviation from the value prescribed by 
the sum rule becomes more and more significant, and the integral finally 
diverges at the instability point.$^{20,21}$
\vspace{\baselineskip} \\
{\bf Spherical model} \\

Another analytical expression for $\alpha({\bf k})$ is given by 
the SM,$^{6,19,22-24}$ also known as the Onsager cavity field
theory,$^{25,26}$
\begin{equation}
\alpha^{SM}({\bf k}) = \frac{1}{ c(1-c) \left[ - \tilde{\Sigma}^{SM} +
2 \beta V({\bf k}) \right] } \ , \label{85}
\end{equation}
where $\tilde{\Sigma}^{SM}$ is, at fixed temperature and concentration, 
a number determined from the sum rule~(\ref{24}). Therefore, the sum 
rule is satisfied in the SM by definition, contrary to the case of the 
RPA. From the definition of the SM it also follows that the self-energy 
depends not only on concentration, like its RPA counterpart~(\ref{73}), 
but also on temperature. Nevertheless, the SM self-energy is still 
wavevector-independent. The explicit expression for $\tilde{\Sigma}^{SM}$ 
can be derived from the sum rule~(\ref{24}):
\begin{eqnarray}
\tilde{\Sigma}^{SM} & = & \tilde{\Sigma}^{RPA} + 
\delta \tilde{\Sigma} \ , \label{124a} \\
\delta \tilde{\Sigma} & = & 2 \beta \sum_{j} \alpha_{ij} V_{ij} = 
\frac{2 \beta}{\Omega} \int d {\bf k} \ \alpha({\bf k}) 
V ({\bf k}) \ . \label{124b} 
\end{eqnarray}
Similarly to the RPA, the SM is analogous to the SSA, since the SM 
self-energy is diagonal in the direct space. However, the SM is rather 
the self-consistent SSA, like the CPA, because the sum rule here plays the 
role of the CPA self-consistency condition. In fact, the sum rule is the 
self-consistency condition: it simply means that the diagonal matrix 
elements of the approximate and exact PCFs are the same. More 
surprisingly, it was shown$^{27,28}$ that both the CPA and the SM 
could be obtained by summation of the same sets of diagrams in the 
corresponding perturbation expansions. We can conclude, therefore, 
that the SM is the CPA for the Ising model.
\vspace{\baselineskip} \\
{\bf Cluster variation method} \\

The CVM$^{29}$ is at present the standard technique for 
quantitative calculation of thermodynamic properties of alloys. It 
is discussed in great detail in almost every book or review on the 
subject.$^{1-3,6}$ We do not attempt to do this here; instead, we will 
consider only those features of the CVM which are relevant to our 
discussion, without going into technical aspects of the method. 

The CVM is essentially a procedure which allows us to derive an 
approximate expression for the variational configurational entropy 
$S(\alpha)$ of the system. The CVM entropy is a function of probabilities 
of various atomic configurations on lattice clusters which belong to the 
so-called basic, or maximal, cluster. A particular CVM approximation for 
$S(\alpha)$ is therefore defined by the choice of the basic cluster. 
Combined with the variational internal energy $E(x,\alpha)$, the CVM 
configurational entropy gives the expression~(\ref{114a}) for the 
variational free energy $\Phi(x,\alpha)$. The operators $a_{n}$ in the 
alloy Hamiltonian are products of the spin variables or of the occupation 
numbers, and the averages $\alpha_{n}$ (Eq.~(\ref{87})) are thus related 
to, or coincide with, cluster probabilities entering the expression 
for the CVM configurational entropy. The variational free energy $\Phi$ 
is then minimized with respect to all cluster probabilities, taking 
into account various self-consistency constraints. 

The self-energy obtained in the framework of the CVM depends, in general, 
on all three parameters -- temperature, concentration and wavevector.
The problem with the CVM, however, is that this method is intrinsically 
numerical and does not lead to analytical approximations for pair 
correlations. The reason for this is that in most cases the number of 
cluster variables used to get a reasonably accurate formula for the 
configurational entropy is far too large. From the point of view of 
the general technique of the Legendre transformations, the CVM 
corresponds to the high-order transformation with respect to all 
coefficients $x_{n}$ conjugated to cluster variables which are 
involved in the expression for the variational configurational 
entropy $S(\alpha)$. Therefore, as far as SRO is concerned, general 
Eq.~(\ref{93}) is still valid, as is the Ising model-specific 
Eq.~(\ref{102}). However, in practical sense this case is very 
different from that of the first Legendre transformation. In the 
latter, the inversion of the matrix 
$\partial^{2} \Phi / \partial \alpha \partial \alpha$ is carried out 
trivially using the Fourier transformation. In the CVM this object in 
the reciprocal space is still a sufficiently large matrix, and in all 
realistic situations it is necessary to resort to numerics. Analytical 
formulae were obtained only for such simple model cases as the pair 
(also known as quasichemical, or Fowler-Guggenheim, or Bethe) 
approximation, or the square (Kramers-Wannier) approximation for the 
nearest-neighbour Ising model on the square lattice.$^{2,20}$
Besides, the CVM suffers from the same drawback as the RPA, though to a 
lesser extent: the integrated intensity is not conserved, and its 
behavior with temperature is similar to the case of the RPA.$^{2,20}$ 
\vspace{\baselineskip} \\
{\bf High-temperature expansion} \\

As we have seen, the RPA and the SM are the analogs of the SSAs 
in the electronic theory of disordered alloys and thus fail to 
take account of the wavevector dependence of the self-energy. 
On the other hand, the CVM leads to the {\bf k}-dependent self-energy, 
but loses the simplicity of the former two approximations by 
not providing analytical expressions for correlations. At the 
semiquantitative level of the RPA this problem can be cured by 
using the HTE for the self-energy.$^{30}$

To do this, we return to Eq.~(\ref{109}) which, in combination 
with Eq.~(\ref{113}), can be written as
\begin{eqnarray}
\Sigma & = & \Sigma^{RPA} + \delta \Sigma \ , \label{116a} \\
 \delta \Sigma_{ij} & = & - \beta \, \frac{\partial^{2} \left( 
\delta \Phi \right)}{\partial m_{i} \partial m_{j}} \ . \label{116b} 
\end{eqnarray}
It is known that the RPA is exact to first order in $1/T$ (i.e., in 
$\Delta$);$^{2}$ it means that $\delta \Phi$ and $\delta \Sigma$ are 
of order $\Delta^{2}$. It is also known how to construct the first 
Legendre transformation $\Gamma$ (and, therefore, $\delta \Phi$) 
for the Ising model.$^{15-17}$ Sorting all contributions to 
$\delta \Phi$ (diagrams) according to the number of lines $\Delta$, 
we obtain the HTE; first eight orders in $\Delta$ are available in 
the literature.$^{17}$ For simplicity reasons, the discussion here 
is confined to two first orders,
\setlength{\unitlength}{2mm}
\thicklines
\begin{equation}
- \beta \, \delta \Phi = 
%
%
\frac{1}{4} 
\begin{picture}(6,3)(0,2.5)
\put(1,3){\circle*{0.5}}
\put(5,3){\circle*{0.5}}
\qbezier(1,3)(3,5)(5,3)
\qbezier(1,3)(3,1)(5,3)
\end{picture}
%
%
+ \frac{1}{12} 
\begin{picture}(6,3)(0,2.5)
\put(1,3){\circle*{0.5}}
\put(5,3){\circle*{0.5}}
\put(1,3){\line(1,0){4}}
\qbezier(1,3)(3,5)(5,3)
\qbezier(1,3)(3,1)(5,3)
\end{picture}
%
%
+ \frac{1}{6} 
\begin{picture}(6,3)(0,2.5)
\put(1,1.85){\circle*{0.5}}
\put(5,1.85){\circle*{0.5}}
\put(3,4.85){\circle*{0.5}}
\put(1,1.85){\line(1,0){4}}
\put(1,1.85){\line(2,3){2}}
\put(5,1.85){\line(-2,3){2}}
\end{picture} 
+ O(\Delta^{4}) \ , \label{117}
\end{equation}
where a line corresponds to $\Delta$, and a vertex with $n$ 
attached lines represents the function $u_{n}(m_{i})$. For 
the diagrams in Eq.~(\ref{117}) $n$ is equal to either 2 or 3:
\begin{eqnarray}
u_{2}(m_{i}) & = & 1-m_{i}^{2} \ , \label{118a} \\
u_{3}(m_{i}) & = & -2 m_{i} \left( 1-m_{i}^{2} \right) \ . \label{118b} 
\end{eqnarray}
The expressions for the diagonal and off-diagonal parts of 
$\delta \Sigma$ thus are, according to Eq.~(\ref{116b}), 
\setlength{\unitlength}{2mm}
\thicklines
\begin{eqnarray}
\delta \Sigma_{ii} & = & 
%
%
\frac{1}{2} 
\begin{picture}(4,3)(0,2.5)
\put(2,1){\circle*{0.5}}
\put(2,5){\circle*{0.5}}
\qbezier(2,1)(0,3)(2,5)
\qbezier(2,1)(4,3)(2,5)
\put(2,1){\line(1,0){1}}
\put(2,1){\line(-1,0){1}}
\end{picture}
%
%
+ \frac{1}{6} 
\begin{picture}(4,3)(0,2.5)
\put(2,1){\circle*{0.5}}
\put(2,5){\circle*{0.5}}
\put(2,1){\line(0,1){4}}
\qbezier(2,1)(0,3)(2,5)
\qbezier(2,1)(4,3)(2,5)
\put(2,1){\line(1,0){1}}
\put(2,1){\line(-1,0){1}}
\end{picture}
%
%
+ \frac{1}{2} 
\begin{picture}(6,3)(0,2.5)
\put(1,4.15){\circle*{0.5}}
\put(5,4.15){\circle*{0.5}}
\put(3,1.15){\circle*{0.5}}
\put(1,4.15){\line(1,0){4}}
\put(1,4.15){\line(2,-3){2}}
\put(5,4.15){\line(-2,-3){2}}
\put(3,1.15){\line(1,0){1}}
\put(3,1.15){\line(-1,0){1}}
\end{picture}
+ O(\Delta^{4}) \ , \label{119a} \\
\delta \Sigma_{ij} & = & 
%
%
\frac{1}{2} 
\begin{picture}(6,3)(0,2.5)
\put(1,3){\circle*{0.5}}
\put(5,3){\circle*{0.5}}
\qbezier(1,3)(3,5)(5,3)
\qbezier(1,3)(3,1)(5,3)
\put(1,3){\line(0,-1){1}}
\put(5,3){\line(0,-1){1}}
\end{picture} 
%
%
+ \frac{1}{6} 
\begin{picture}(6,3)(0,2.5)
\put(1,3){\circle*{0.5}}
\put(5,3){\circle*{0.5}}
\put(1,3){\line(1,0){4}}
\qbezier(1,3)(3,5)(5,3)
\qbezier(1,3)(3,1)(5,3)
\put(1,3){\line(0,-1){1}}
\put(5,3){\line(0,-1){1}}
\end{picture}
%
%
+ \begin{picture}(6,3)(0,2.5)
\put(1,1.85){\circle*{0.5}}
\put(5,1.85){\circle*{0.5}}
\put(3,4.85){\circle*{0.5}}
\put(1,1.85){\line(1,0){4}}
\put(1,1.85){\line(2,3){2}}
\put(5,1.85){\line(-2,3){2}}
\put(1,1.85){\line(0,-1){1}}
\put(5,1.85){\line(0,-1){1}}
\end{picture} 
+ O(\Delta^{4}) \ , \ \ i \neq j \ . \label{119b} 
\end{eqnarray}
Here a vertex with $n$ internal lines and $k$ external legs
corresponds to the $k$th derivative of the function $u_{n}(m)$.
The corresponding analytical expressions have the form
\begin{eqnarray}
\delta \Sigma_{ii} & = & - \sum_{l} (1-m_{l}^{2}) \Delta_{il}^{2} 
- 4 m_{i} \sum_{l} m_{l} (1-m_{l}^{2}) \Delta_{il}^{3} \nonumber \\
& & - \sum_{kl} (1-m_{k}^{2}) (1-m_{l}^{2}) \Delta_{ik} \Delta_{il} 
\Delta_{kl} + O(\Delta^{4}) \ , \label{120a} \\
\delta \Sigma_{ij} & = & 2 m_{i} m_{j} \Delta_{ij}^{2} 
+ \frac{2}{3} (1-3m_{i}^{2}) (1-3m_{j}^{2}) \Delta_{ij}^{3} \nonumber \\
& & + 4 m_{i} m_{j} \Delta_{ij} \sum_{l} (1-m_{l}^{2}) 
\Delta_{il} \Delta_{jl} + O(\Delta^{4}) \ , \ \ i \neq j \ . \label{120b} 
\end{eqnarray}
In the homogeneous case $m_{i}=m$ for all sites $i$. Defining constants
\begin{equation}
a_{1} = \sum_{l} J_{il}^{2} \ , \ \ \ 
a_{2} = \sum_{l} J_{il}^{3} \ , \ \ \ 
a_{3} = \sum_{kl} J_{ik} J_{il} J_{kl}  \label{121}
\end{equation}
and lattice functions
\begin{equation}
(f_{1})_{ij} = J_{ij}^{2} \ , \ \ \
(f_{2})_{ij} = J_{ij}^{3} \ , \ \ \
(f_{3})_{ij} = J_{ij} \sum_{l} J_{il} J_{jl} \ , \label{122}
\end{equation}
which depend only on the interaction $J$, one can finally write 
the expression for $\delta \Sigma$ in the ${\bf k}$-space:
\begin{eqnarray}
\delta \Sigma({\bf k}) & = & - (1-m^{2}) \beta^{2} a_{1} 
+ 2 m^{2} \beta^{2} f_{1}({\bf k}) 
- 4 m^{2} (1-m^{2}) \beta^{3} a_{2} \nonumber \\
& & - (1-m^{2})^{2} \beta^{3} a_{3} 
+ \frac{2}{3} (1-3m^{2})^{2} \beta^{3} f_{2}({\bf k}) \nonumber \\ 
& & + 4 m^{2} (1-m^{2}) \beta^{3} f_{3}({\bf k}) + O(\beta^{4}) \ .
\label{123a}
\end{eqnarray}
In alloy notations this result reads
\begin{eqnarray}
\delta \tilde{\Sigma}({\bf k}) & = & - 4 c(1-c) \beta^{2} \tilde{a}_{1} 
+ 2 (1-2c)^{2} \beta^{2} \tilde{f}_{1}({\bf k}) 
+ 8 c(1-c) (1-2c)^{2} \beta^{3} \tilde{a}_{2} \nonumber \\
& & + 8 c^{2} (1-c)^{2} \beta^{3} \tilde{a}_{3} 
- \frac{4}{3} [1-6c(1-c)]^{2} \beta^{3} \tilde{f}_{2}({\bf k}) \nonumber \\
& & - 8 c(1-c) (1-2c)^{2} \beta^{3} \tilde{f}_{3}({\bf k}) + O(\beta^{4}) \ ,
\label{123b}
\end{eqnarray}
where
\begin{eqnarray}
\tilde{a}_{1} & = & \sum_{l} V_{il}^{2} \ , \ \ \
\tilde{a}_{2} = \sum_{l} V_{il}^{3} \ , \ \ \
\tilde{a}_{3} = \sum_{kl} V_{ik} V_{il} V_{kl} \ , \label{123c} \\
(\tilde{f}_{1})_{ij} & = & V_{ij}^{2} \ , \ \ \
(\tilde{f}_{2})_{ij} = V_{ij}^{3} \ , \ \ \
(\tilde{f}_{3})_{ij} = V_{ij} \sum_{l} V_{il} V_{jl} \ . \label{123d}
\end{eqnarray}
\vspace{\baselineskip} \\
{\bf Alpha- and gamma-expansions} \\

The approximate expressions~(\ref{123a}), (\ref{123b}) for the 
self-energy given by the first orders of the HTE satisfy almost all 
requirements: they are analytical, relatively simple 
and take into account dependence on all relevant parameters, 
including the wavevector. However, the limits of applicability of 
the HTE are essentially the same as those of the RPA; the HTE is 
quantitatively correct only at reasonably high temperatures, as can 
be concluded already from the name of the expansion. What is needed, 
therefore, is an approximation which would combine all the 
advantages of the HTE self-energy with the applicability at moderate 
temperatures, including the range not far away from the transition 
or instability points. 

The theory of SRO which will now be discussed$^{12,13}$ is based 
on the fairly general procedure$^{27}$ of self-consistent renormalization 
of the bare propagator $\Delta^{-1}$ in the functional-integral 
representation of the generating functional for correlation functions. 
The resulting expansion for the matrix elements of the self-energy is in 
powers of the matrix elements of the fully dressed propagator, which 
in the case of the Ising model coincides with the PCF~(\ref{65}). 
Two first non-zero orders of this expansion for the off-diagonal part 
of the self-energy were calculated,$^{12,13}$
\begin{eqnarray}
\Sigma_{ij} & = & a G_{ij}^{2} + 
b G_{ij}^{3} + O(G^{4}) \ , \ \ i \neq j \ , \label{133a} \\
a & = & \frac{2m^{2}}{(1-m^{2})^{4}} \ , \label{133b} \\
b & = & \frac{2[(1-3m^{2})^{2}-12m^{4}]}{3(1-m^{2})^{6}} \ , \label{133c}
\end{eqnarray}
in terms of the alloy variables
\begin{eqnarray}
\tilde{\Sigma}_{ij} & = & \tilde{a} \alpha_{ij}^{2} + \tilde{b} 
\alpha_{ij}^{3} + O(\alpha^{4}) \ , \ \ i \neq j \ , \label{134a} \\
\tilde{a} & = & \frac{(1-2c)^{2}}{2[c(1-c)]^{2}} \ , \label{134b} \\
\tilde{b} & = & \frac{[1-6c(1-c)]^{2}-3(1-2c)^{4}}{6[c(1-c)]^{3}} 
\ . \label{134c}
\end{eqnarray}
Eq.~(\ref{134a}) is the expansion in powers of the SRO parameters, and is 
therefore referred to as the AE,$^{31}$ though initially it was obtained 
in the framework of the GEM.$^{12,13}$ The difference between the 
AE and GEM is discussed below. In the two calculated orders the AE 
preserves the sum rule~(\ref{24}), and the expression for the diagonal 
part of the self-energy then comes from Eqs.~(\ref{24}) and (\ref{134a}), 
similarly to the case of the SM (Eqs.~(\ref{124a}), (\ref{124b})): 
\begin{eqnarray}
\tilde{\Sigma}_{ii} & = & \tilde{\Sigma}^{RPA} +
\delta \tilde{\Sigma}_{ii} \ , \label{135a} \\
\delta \tilde{\Sigma}_{ii} & = & 2 \beta \sum_{j} \alpha_{ij} V_{ij} -
\sum_{j(\neq i)} \left( \tilde{a} \alpha_{ij}^{3} + 
\tilde{b} \alpha_{ij}^{4} \right) + O(\alpha^{5}) \ . \label{135b}
\end{eqnarray}
Note that the first term in Eq.~(\ref{135b}) corresponds to the SM 
(cf. Eq.~(\ref{124b})) which is the zero-order approximation for 
both the AE and the GEM; in the SM the self-energy is diagonal 
($\tilde{a}=\tilde{b} =0$). 

The difference between the GEM and the AE is in the way of selection 
of leading terms in the expansion~(\ref{134a}). The GEM, originally 
proposed by Tokar$^{27}$ and further developed by Tokar, Grishchenko and 
the author,$^{12,13,32-35}$ is based on the assumption that correlations 
decrease rapidly with distance, and the GEM expansion parameter is 
\begin{equation}
\gamma = \exp (- 1 / \xi) \ , \label{136}
\end{equation}
where $\xi$ is the dimensionless correlation length. The leading terms 
in the diagrammatic expansion for the self-energy are selected in the 
framework of the GEM according to the total length of all lines in the 
diagrams, where the line connecting lattice sites $i$ and $j$ represents 
$\alpha_{ij}$. For example, in the case of three Bravais lattices 
belonging to the cubic system taking into account several first terms
of the perturbation theory leads to the result$^{12,13}$
\begin{eqnarray}
\tilde{\Sigma}_{s} & = & \tilde{a} \alpha_{s}^{2} + 
\tilde{b} \alpha_{s}^{3} \ , \ \ s = 1 \ , \label{137a} \\
\tilde{\Sigma}_{s} & = & \tilde{a} \alpha_{s}^{2} \ , 
\ \ s = 2,3 \ , \label{137b} \\ 
\tilde{\Sigma}_{s} & = & 0 \ , \ \ s > 3 \ , \label{137c}
\end{eqnarray}
where subscript $s$ denotes the matrix elements corresponding to the $s$th
coordination shell. However, the GEM assumption about the rapid decay of 
correlations is not always valid; e.g., it is incorrect in the cases where 
distant interactions 
are essential. The AE abandons this assumption and uses instead the SRO 
parameters $\alpha_{ij}$ themselves as the expansion parameters; the 
leading terms are chosen according to the number of lines in the diagrams 
(i.e., the powers of $\alpha_{ij}$), since all $\alpha_{ij}$ are relatively 
small. A particular AE approximation is defined by neglecting higher-order 
terms and including only finite number of coordination shells in the AE 
expression~(\ref{134a}) for the off-diagonal part of the self-energy. 
The GEM was successfully applied to both the direct and inverse problems 
of alloy diffuse scattering,$^{12,13,36-38}$ leading to reliable 
results everywhere except in the vicinity of the instability point, 
while the AE was used in the analysis of some of the diffuse-scattering 
anomalies discussed in the next section.$^{31,39}$ 
\vspace{2 \baselineskip} \\

%% file: part3.tex
{\bf ANOMALIES IN ALLOY DIFFUSE SCATTERING} \\

In this last section we will show how expression~(\ref{70}) for the 
intensity $\alpha({\bf k})$ and, in particular, the wavevector dependence 
of the self-energy $\tilde{\Sigma}({\bf k})$ lead to straightforward 
explanation of recently observed unusual features (anomalies) of diffuse 
scattering from disordered alloys and to prediction of some new effects. 
\vspace{\baselineskip} \\
{\bf Temperature-dependent Fermi surface-induced peak splitting} \\ 

This curious effect (the temperature dependence of the splitting) was 
discovered in 1996 by Reichert, Moss and Liang$^{40}$ in the first 
{\em in situ} experiment to resolve the fine structure of the 
equilibrium diffuse scattering intensity from the disordered Cu$_{3}$Au 
alloy. The separation of the split maxima changed reversibly, increasing 
with temperature. The same behavior of the splitting was also found$^{41}$ 
by analysing results of the MC simulations for the Cu$_{0.856}$Al$_{0.144}$ 
alloy.$^{42}$ The fourfold splitting of the intensity peaks located at the 
(110) and equivalent positions (Figure~1) is attributed to the indirect
interaction of atoms via conduction electrons in an alloy whose Fermi
surface has flat portions; the effective interatomic pair interaction
$V({\bf k})$ itself has split minima in the reciprocal space, and 
their location is determined by the wavevector $2 {\bf k}_{F}$ 
spanning these flat portions of the Fermi surface.$^{5,43}$ It is 
usually assumed that $V({\bf k})$ is temperature-independent. This 
assumption is justified at least as far as positions of the $V({\bf k})$ 
minima are concerned, since the $2 {\bf k}_{F}$ value is unlikely to 
change over the considered temperature range. Besides, the MC
calculations$^{42}$ in which the increase of the splitting
with temperature was found$^{41}$ were carried out for the
temperature-independent pair interaction parameters. The standard 
RPA (KCM) treatment (Eq.~(\ref{72})) predicts that positions of the 
intensity peaks coincide with those of the corresponding minima of 
$V({\bf k})$, and, therefore, the splitting does not depend on temperature. 
\vspace{\baselineskip} \\
\begin{figure}[h]
\begin{center}
\includegraphics[angle=0]{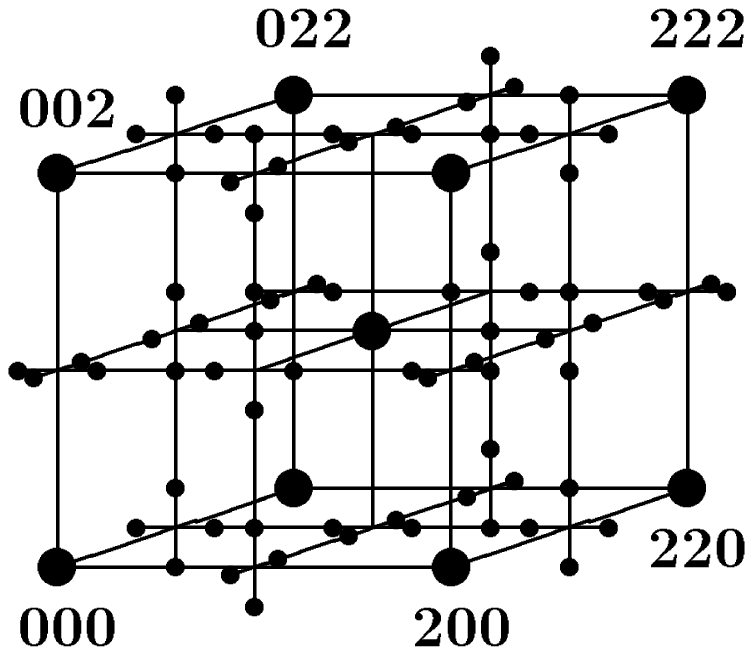} 
\end{center}
\end{figure}
\vspace{0.01 \baselineskip} \\
{\footnotesize {\bf Figure 1.} 3D reciprocal-space picture of scattering 
from the FCC alloys discussed in the text. Large dots represent the Bragg 
reflections. Characteristic crosses formed by small dots correspond to 
the split diffuse intensity peaks.}
\vspace{\baselineskip} \\

As will be demonstrated below, this phenomenon can be easily understood 
by employing the notion of the {\bf k}-dependent self-energy.$^{31}$ 
Let us consider the $\alpha({\bf k})$ profile along one of the lines 
containing split peaks, e.g., the (h10) line, and concentrate on
the two peaks around the (110) position. The peak positions $k_{\alpha}$
($k$ is the deviation of the wavevector from the (110) position 
along the (h10) line) are determined by the condition 
$\partial \alpha / \partial k = 0$ which gives
\begin{equation}
2 \, \frac{\partial V}{\partial k} = 
T \, \frac{\partial \tilde{\Sigma}}{\partial k} \ . \label{125}
\end{equation}
Eq.~(\ref{125}) means that the ${\bf k}$-dependence of $\tilde{\Sigma}$ 
leads to the shift $\delta k = k_{\alpha} - k_{V}$ of the peak position 
with respect to the position $k_{V}$ of the corresponding minimum of 
$V({\bf k})$ (the latter is defined by the condition 
$\partial V / \partial k = 0$). Furthermore, the right-hand side of 
Eq.~(\ref{125}) is a function of $T$, while its left-hand side is 
$T$-independent. The $\alpha({\bf k})$ peaks will therefore change 
their positions with temperature. The ``local'' 
temperature behavior of the splitting is reflected in the sign of the 
derivative $\partial_{T} k_{I}$ which can be calculated by expanding
the derivatives in Eq.~(\ref{125}) in powers of small changes of $T$ 
and $k_{I}$ and retaining only linear terms:
\begin{equation}
\frac{\partial k_{\alpha}}{\partial T} =
\left[ \left( \frac{\partial \tilde{\Sigma}}{\partial k}
+ T \, \frac{\partial^{2} \tilde{\Sigma}}{\partial k \partial T} \right)
\left/ \left( 2 \, \frac{\partial^{2} V}{\partial k^{2}} - T \,
\frac{\partial^{2} \tilde{\Sigma}}{\partial k^{2}} \right) \right.
\right]_{k=k_{\alpha}} \, . \label{126}
\end{equation}
Our aim now is to develop a kind of minimal, i.e., simplest possible, 
theory which would be able to describe essential features of the 
considered effect (and, apropos, two other anomalies discussed in this 
section). Interestingly, the approach formally rather similar 
to the Landau theory of second-order phase transitions$^{7}$ could be 
used. Indeed, in the case of not very large splitting the expansion of 
$V(k)$ and $\tilde{\Sigma}(k)$ in powers of $k$ can be used. 
To describe the split minimum of $V(k)$, only the second- and 
fourth-order terms are necessary; since the (110) position serves as 
the origin, the expansions do not contain odd powers of $k$. 
We therefore assume that in the area of the splitting $V(k)$ and 
$\tilde{\Sigma}(k)$ have the following approximate form, 
\begin{eqnarray}
V(k) & = & V(0) + \frac{1}{2} A_{V} k^{2} + 
\frac{1}{4} B_{V} k^{4} \ , \label{127a} \\
\tilde{\Sigma}(k) & = & \tilde{\Sigma}(0) + \frac{1}{2} 
A_{\Sigma} k^{2} + \frac{1}{4} B_{\Sigma} k^{4} \ , \label{127b} 
\end{eqnarray}
where $A_{V}<0$, $B_{V}>0$, the signs of $A_{\Sigma}$ and $B_{\Sigma}$ are 
arbitrary (there are no apparent restrictions on the behavior of the 
self-energy), and $k=0$ corresponds to the (110) position. The resulting 
inverse intensity $\alpha^{-1}(k)$ has exactly the form of the Landau free 
energy functional in the low-symmetry phase where the latter possesses a double 
minimum; this implies $2|A_{V}|+TA_{\Sigma}>0$ and $2B_{V}-TB_{\Sigma}>0$. 
The wavevector $k$ plays the role of the order parameter. Substituting 
approximations (\ref{127a}) and (\ref{127b}) into general Eqs.~(\ref{125}) 
and (\ref{126}), we get
\begin{eqnarray}
k_{\alpha} & = & \pm \sqrt{\frac{2 |A_{V}| +
T A_{\Sigma}}{2 B_{V} - T B_{\Sigma}}} \ , \label{128a} \\
k_{\alpha}^{-1} \frac{\partial k_{\alpha}}{\partial T} & = & 
\frac{1}{2} \left[ \frac{A_{\Sigma} + T \, \partial A_{\Sigma} / 
\partial T}{2 |A_{V}| + T A_{\Sigma}} + 
\frac{B_{\Sigma} + T \, \partial B_{\Sigma} / \partial T}{
2 B_{V} - T B_{\Sigma}} \right] \ , \label{128b} 
\end{eqnarray}
while $k_{V}=\pm \sqrt{|A_{V}|/B_{V}}$. It is seen that the shifts of the
two peaks and their temperature derivatives have opposite signs and the
same absolute values. Eq.~(\ref{128b}) clearly shows two scenarios for 
the temperature behavior of the splitting, depending on the sign of its 
right-hand side which can be either positive or negative. The first one 
is the increase of the splitting with temperature discussed above; this 
takes place when the right-hand side of Eq.~(\ref{128b}) is negative. 
Apart from that, the theory predicts that the decrease of the splitting 
with increasing temperature is also possible. This regime corresponds to 
the case of positive right-hand side of Eq.~(\ref{128b}), and such 
temperature dependence has not yet been observed experimentally. Thus, 
the behavior of the self-energy determines whether the splitting increases 
or decreases with temperature. At high temperatures the correction 
$\delta \Sigma$ to the wavevector-independent $\Sigma^{RPA}$ 
(Eq.~(\ref{116a})), and, therefore, $A_{\Sigma}$ and $B_{\Sigma}$, 
are of order $\Delta^{2}$ 
(i.e., $T^{-2}$). From Eq.~(\ref{128a}) it then follows that the absolute 
value of the shift $\delta k$ decreases as $T^{-1}$ with temperature, unless 
the corresponding coefficient identically vanishes. 
\vspace{\baselineskip} \\
{\bf Coalescence of Fermi surface-related intensity peaks} \\

The analogy with the Landau theory of phase transitions, though rather 
formal, immediately leads to the following question: in the considered 
case, what would correspond to the transition point? The answer to this 
question is fairly obvious; there exists a possibility for the splitting 
to disappear at some point as temperature decreases, before the 
transition to the low-symmetry phase occurs.$^{44}$ This anomaly was 
neither observed experimentally nor correctly described theoretically, 
though the coalescence of intensity peaks (unrelated to any Fermi 
surface effects) was found for the exactly solvable 1D Ising model$^{45}$
and in the CVM calculations for the 2D ANNNI model.$^{46}$ As we will 
show in this subsection, Eq.~(\ref{70}) provides clear understanding 
of how such effect takes place.

In the treatment given above it is, in fact, implicitly assumed that 
the wavevector dependence of the interaction term $2 \beta V({\bf k})$ 
in Eq.~(\ref{70}) dominates, at least in the area of the splitting, 
i.e., near the (110) position. In this case the shape of the diffuse 
intensity closely follows that of $V({\bf k})$; in particular, there 
exists one-to-one correspondence between the split minima of the 
interaction and the split intensity peaks. The variation of the 
self-energy with ${\bf k}$ in this part of the reciprocal space is 
comparatively weak, though qualitatively important for the description 
of the temperature-dependent splitting. This assumption is certainly correct
at sufficiently high temperatures, where the RPA (KCM) approximation in
which the self-energy is ${\bf k}$-independent works reasonably well.
Meanwhile, as temperature starts to decrease, the correction to the RPA 
self-energy ($\propto T^{-2}$) grows faster than the interaction term 
$2 \beta V({\bf k})$ ($\propto T^{-1}$). We can then encounter a situation 
when the variations of ${\tilde\Sigma}({\bf k})$ and $2 \beta V({\bf k})$ 
with the wavevector are of the same order of magnitude. With temperature 
further decreasing, the wavevector dependence of the self-energy can even 
become dominant. In this regime positions of the diffuse intensity peaks 
would be determined by the maxima of $\tilde{\Sigma}({\bf k})$. 
\vspace{\baselineskip} \\
\begin{figure}[h]
\begin{center}
\includegraphics[angle=-90]{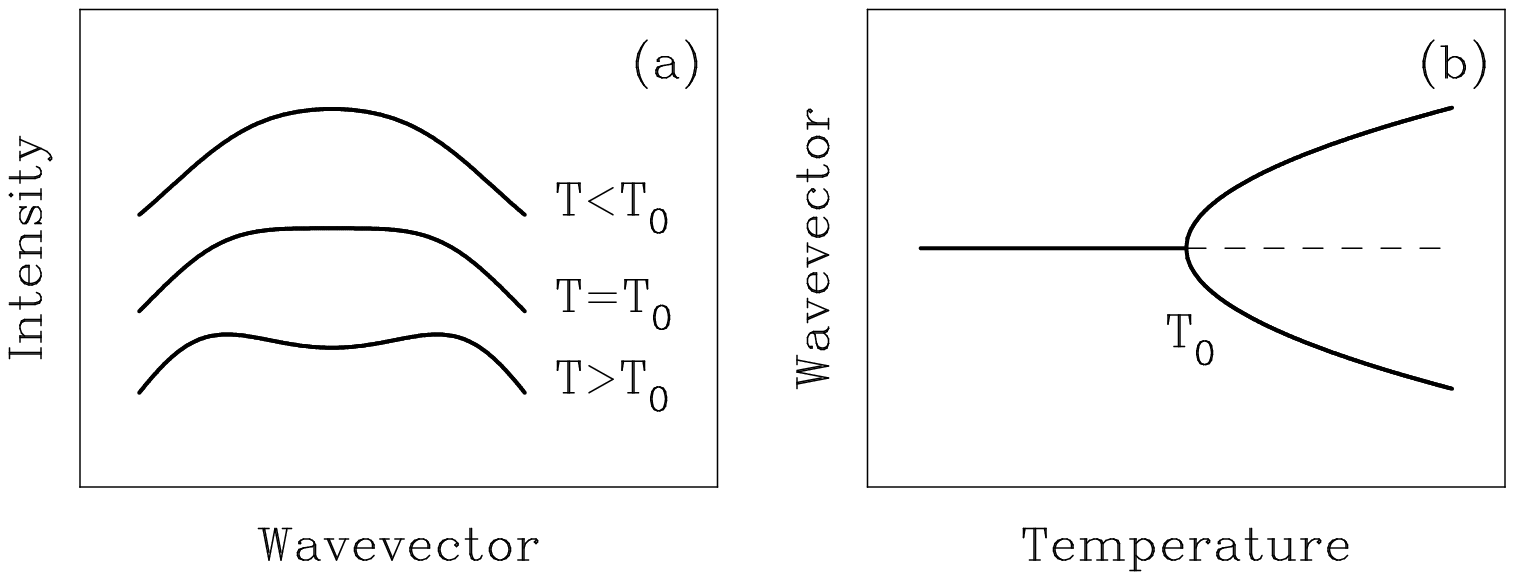}
\end{center}
\end{figure}
\vspace{-0.5 \baselineskip} \\
{\footnotesize {\bf Figure 2.} Schematical temperature dependence of the 
intensity profile (a) and the peak positions (b) in the coalescence case.}
\vspace{\baselineskip} \\

The behavior of the self-energy in the ${\bf k}$-space is, in general, 
qualitatively different from that of the interaction. In particular, 
there is no special reason to expect that the self-energy would have any 
extrema away from the special points. Let us assume that the self-energy does 
not have such extrema and that the variation of the self-energy in the 
reciprocal space becomes more and more important in comparison with the 
corresponding variation of $2 \beta V({\bf k})$ as temperature decreases. 
Then the qualitative picture of the temperature behavior of the splitting 
is as follows. At high temperatures the self-energy is almost 
${\bf k}$-independent, and the intensity peak positions deviate little 
from those of the corresponding minima of $V({\bf k})$. As temperature 
decreases, the wavevector dependence of $\tilde{\Sigma}$ becomes more 
pronounced; the peak positions move farther away from the positions of 
$V({\bf k})$ minima and towards those special points ${\bf k}_{0}$ at which 
$\tilde{\Sigma}({\bf k})$ has maxima. Eventually, as temperature reaches 
certain value $T_{0}$, the intensity peaks coalesce at these special points
and the splitting disappears (Figure~2).

The coalescence temperature $T_{0}$ can be found from the condition of
vanishing second derivative
$\alpha^{\prime \prime} = \partial^{2} \alpha / \partial k^{2}$
of the intensity with respect to the wavevector at $ k_{0}$;
the sign of this derivative controls the presence or absence of the 
splitting. At the special points all the first derivatives are equal 
to zero, and from Eq.~(\ref{70}) we obtain
\begin{equation}
\alpha^{\prime \prime}(k_{0}) = c(1-c) \alpha^{2}(k_{0})
\left[ \tilde{\Sigma}^{\prime \prime}(k_{0}) -
2 \beta V^{\prime \prime}(k_{0}) \right] \ . \label{129}
\end{equation}
Therefore, the splitting disappears when the second derivatives, 
or curvatures, of the self-energy and of the interaction term 
$2 \beta V(k)$ at the special point $k_{0}$ compensate 
each other, i.e., when
\begin{equation}
\tilde{\Sigma}^{\prime \prime}(k_{0}) =
2 \beta V^{\prime \prime}(k_{0}) \ . \label{130}
\end{equation}

To analyse the behavior of the splitting close to the coalescence point, 
it is convenient to use the same Landau-type approach as in the previous 
case. In this temperature range the splitting above $T_{0}$ is small, and 
expansions~(\ref{127a}) and (\ref{127b}) are valid. Substituting them 
into Eq.~(\ref{70}), 
we get
\begin{equation}
\alpha^{-1}(k) = \alpha^{-1}(0) + \frac{1}{2} A k^{2} +
\frac{1}{4} B k^{4} \ , \label{131} \\
\end{equation}
where second-order coefficient 
\begin{equation}
A = c(1-c) (-A_{\Sigma} + 2 \beta A_{V}) \label{132} \\
\end{equation}
vanishes at $T=T_{0}$ (see Eq.~(\ref{130})), while the fourth-order
coefficient $B$ remains positive at that temperature. We can then,
at temperatures close to $T_{0}$, regard $A$ as linear in $T-T_{0}$ 
with a negative coefficient and $B$ as temperature-independent. 
Thus, the inverse intensity 
$\alpha^{-1}(k)$ behaves almost in the same way that the Landau free 
energy. The only difference here is that the role of temperature is 
reversed; $\alpha^{-1}(k)$ has a double minimum above the coalescence
temperature $T_{0}$ and a single minimum below it. Therefore, at small
positive values of $T-T_{0}$ the splitting increases with temperature
as $(T-T_{0})^{1/2}$. Contrary to the corresponding result of the genuine 
Landau theory, obtained bifurcation exponent is exact, since the 
intensity is an analytical function of the wavevector and can legitimately 
be expanded into the Taylor series. At higher temperatures 
behavior of the splitting changes, and sufficiently far away from 
$T_{0}$ it starts to approach the value dictated by the interaction 
$V(k)$. 
\vspace{\baselineskip} \\
{\bf ``Thermal'' splitting of intensity peaks } \\

We have just considered the situation when the double-well profile of 
the interaction in the vicinity of a special point is compensated by 
the wavevector dependence of the self-energy which has a simple maximum 
at this position. The competing curvatures 
$2 \beta V^{\prime \prime}(k_{0})$ and 
$\tilde{\Sigma}^{\prime \prime}(k_{0})$ are both negative. As a 
result, the second derivative~(\ref{129}) of the intensity vanishes at 
some temperature, and the splitting induced by the Fermi surface effects 
disappears.

It is not very difficult to realize that another kind of curvature 
compensation is possible; this is the case in which both curvatures are 
positive.$^{39}$ This is, perhaps, the most physically interesting situation: 
here the interaction with a single minimum produces the intensity peak with 
no fine structure at higher temperatures (in full accordance with the 
RPA-like considerations), but then, as temperature decreases and the 
wavevector dependence of the self-energy becomes more and more significant,
the compensation takes place and the intensity peak splits. This is 
especially probable when the minimum of $V({k})$ is shallow (i.e., 
$V^{\prime \prime}(k_{0})$ is small); in particular, in the limiting case of 
vanishing $V^{\prime \prime}(k_{0})$ it is the curvature of 
$\tilde{\Sigma}({k})$ 
that controls the fine structure (single- vs. double-peak) of the 
maximum of $\alpha(k)$. The application of the Landau-type description 
gives essentially the same results as before. However, the coefficient 
in front of $T-T_{0}$ in A is now positive; the inverse intensity 
$\alpha^{-1}(k)$ has a single minimum above the splitting temperature 
$T_{0}$ and a double minimum below this point, and the splitting increases 
as $(T_{0}-T)^{1/2}$ with decreasing temperature at small negative 
$T-T_{0}$ values (Figure~3) . The bifurcation exponent is again exact, 
for the reasons mentioned above.
\vspace{\baselineskip} \\
\begin{figure}[h]
\begin{center}
\includegraphics[angle=-90]{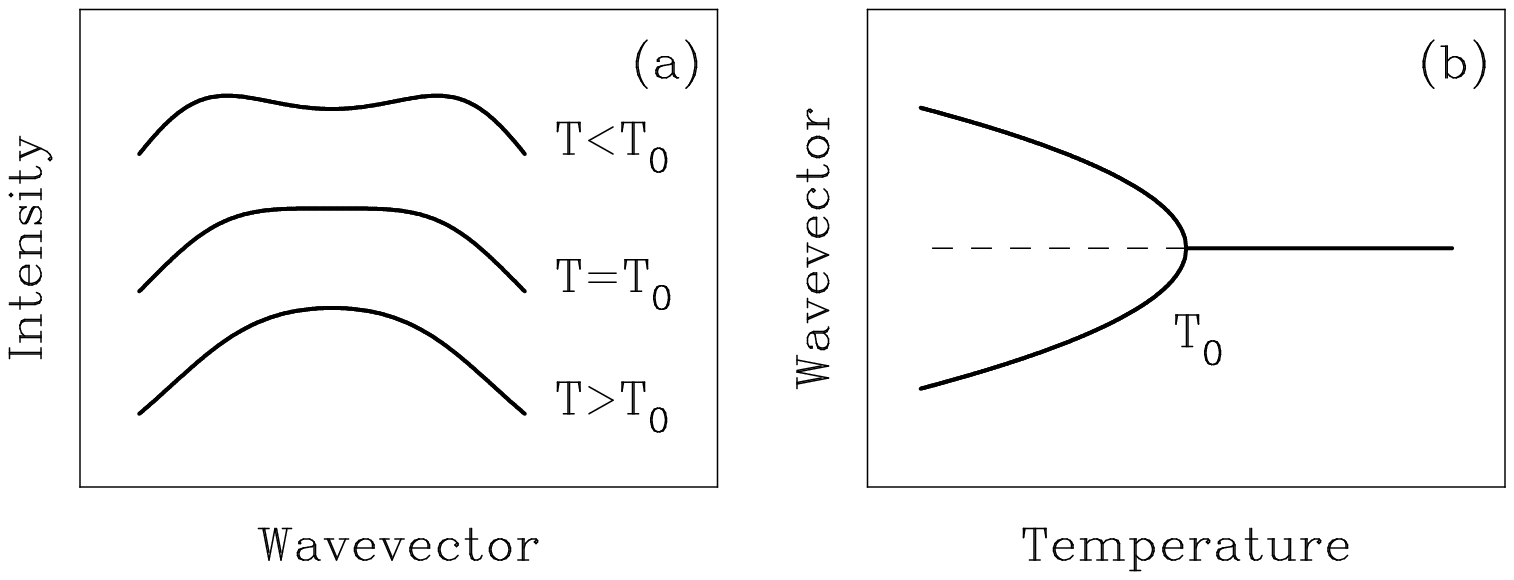}
\end{center}
\end{figure}
\vspace{-0.5 \baselineskip} \\
{\footnotesize {\bf Figure 3.} Schematical temperature dependence of the 
intensity profile (a) and the peak positions (b) in the case of the 
``thermal'' splitting.}
\vspace{\baselineskip} \\

This type of behavior was recently experimentally observed (and 
subsequently reproduced in the MC simulations) for the Pt-V alloy 
system by Le Bolloc'h et al.$^{37}$ In this system 
the splitting of the (100) intensity peak along the (h00) line 
occured with decreasing concentration of vanadium rather then temperature. 
The explanation of this anomaly was proposed by 
the author,$^{39}$ who also pointed out that similar splitting should take 
place when temperature decreases at fixed composition. The predicted 
effect was then discovered in the MC simulations by the same group.$^{47}$
Experimental confirmation of its existence remains to be seen.
\vspace{2 \baselineskip} \\
{\bf REFERENCES} 
\vspace{\baselineskip} \\
1. D. de Fontaine, {\em Solid State Physics} 34:73 (1979); 
47:33 (1994). \\
2. F. Ducastelle. {\em Order and Phase Stability 
in Alloys}, North-Holland, Amsterdam \\ 
\hspace*{13mm} (1991). \\
3. G. Inden and W. Pitsch, in: {\em Phase Transformations 
in Materials}, P.~Haasen, ed., \\
\hspace*{13mm} VCH Press, New York, (1991). \\
4. A.G. Khachaturyan. {\em Theory of Structural 
Transformations in Solids}, Wiley, New \\
\hspace*{13mm} York, (1983). \\
5. M. A. Krivoglaz. {\em Theory of X-Ray and Thermal 
Neutron Scattering by Real Crystals}, \\ 
\hspace*{13mm} Plenum, New York (1969); {\em Diffuse 
Scattering of X-Rays and Neutrons by \\
\hspace*{13mm} Fluctuations}, Springer, Berlin (1996). \\
6. J. M. Ziman. {\em Models of Disorder}, Cambridge University Press,
Cambridge (1979). \\
7. L.D. Landau and E.M. Lifshitz. {\em Statistical Physics}, 
Part~I, Pergamon, Oxford (1980). \\
8. Yu.A. Izyumov and Yu.N. Skryabin. {\em Statistical 
Mechanics of Magnetically Ordered \\
\hspace*{13mm} Systems}, Consultants Bureau, 
New York and London (1988). \\
9. H. Ehrenreich and L.M. Schwartz, {\em Solid State Phys.} 
31:149 (1976). \\
10. R.J. Elliott, J.A. Krumhansl, and P.L. Leath, 
{\em Rev. Mod. Phys.} 46:465 (1974). \\
11. J.S. Faulkner, {\em Prog. Mater. Sci.} 27:1 (1982). \\
12. V.I. Tokar, I.V. Masanskii, and T.A. Grishchenko,
{\em J. Phys. Condens. Matter} \\
\hspace*{13mm} 2:10199 (1990). \\
13. I.V. Masanskii, V.I. Tokar, and T.A. Grishchenko, 
{\em Phys. Rev. B} 44:4647 (1991). \\
14. D.J. Amit. {\em Field Theory, the Renormalization 
Group and Critical Phenomena}, \\ 
\hspace*{13mm} World Scientific, Singapore (1984). \\
15. A.N. Vassiliev. {\em Functional Methods in Quantum Field Theory
and Statistics}, \\ 
\hspace*{13mm} Leningrad State University Press, Leningrad (1976). \\
16. A.N. Vassiliev and R.A. Radzhabov, {\em Teor. Mat. Fiz.} 
21:49 (1974); 23:366 (1975). \\
17. N.M. Bogoliubov et al., {\em Teor. Mat. Fiz.}  26:341 (1975). \\
18. P.C. Clapp and S.C. Moss, {\em Phys. Rev.} 142:418 (1966);
171:754 (1968). \\
19. R. Brout. {\em Phase Transitions}, Benjamin, New York (1965). \\
20. J. M. Sanchez, {\em Physica A} 111:200 (1982). \\
21. T. Mohri, J.M. Sanchez and D. de Fontaine, {\em Acta Metall.}
33:1463 (1985). \\
22. G.S. Joyce, in: {\em Phase Transitions 
and Critical Phenomena}, Vol.~2, C.~Domb and \\
\hspace*{13mm} M.S.~Green, eds., Academic Press, New York (1972). \\
23. J. Philhours and G.L. Hall, {\em Phys. Rev.} 170:496 (1968). \\
24. D.W. Hoffmann, {\em Metall. Trans.} 3:3231 (1972). \\
25. L. Onsager, {\em J. Am. Chem. Soc.} 58:1468 (1936). \\
26. J.B. Staunton and B.L. Gyorffy, {\em Phys. Rev. Lett.} 
69:371 (1992). \\
27. V.I. Tokar, {\em Phys. Lett. A} 110:453 (1985). \\
28. V. Jani\v{s}, {\em Czech. J. Phys. B} 36:1107 (1986); 
{\em Phys. Rev. B} 40:11331 (1989). \\
29. R. Kikuchi, {\em Phys. Rev.} 81:988 (1951). \\
30. I. Tsatskis (unpublished). \\
31. I. Tsatskis (submitted). \\
32. V.I. Tokar and I.V. Masanskii, {\em Fiz. Metal. Metalloved.}
64:1207 (1987). \\
33. I.V. Masanskii and V.I. Tokar, {\em Teor. Mat. Fiz.} 76:118 (1988). \\
34. T.A. Grishchenko, I.V. Masanskii, and V.I. Tokar,
{\em J. Phys. Condens. Matter} \\ 
\hspace*{13mm} 2:4769 (1990). \\
35. I.V. Masanskii and V.I. Tokar, {\em J. Phys. I France} 
2:1559 (1992). \\
36. L. Reinhard and S.C. Moss, {\em Ultramicroscopy} 52:223 (1993). \\
37. D. Le Bolloc'h et al., in: {\em Proceedings of 
the Joint NSF/CNRS Workshop on Alloy \\
\hspace*{13mm} 
Theory}, Mont Sainte Odile Monastery, 
Strasbourg, France, October 11-15, 1996, \\ 
\hspace*{13mm} 
{\em Comput. Mater. Sci.} 8:24 (1997). \\
38. M. Borici-Kuqo and R. Monnier, Ref.~37, p.~16. \\
39. I. Tsatskis (submitted). \\
40. H. Reichert, S.C. Moss and K.S. Liang, {\em Phys. Rev. Lett.} 
77:4382 (1996). \\
41. S.C. Moss and H. Reichert (private communication);
H. Reichert, I. Tsatskis and \\ 
\hspace*{13mm} S.C. Moss, Ref.~37, p.~46. \\
42. H. Roelofs et al., {\em Scripta Mat.} 34:1393 (1996). \\
43. S.C. Moss, {\em Phys. Rev. Lett.} 22:1108 (1969);
S.C. Moss and R.H. Walker, {\em J. Appl. \\
\hspace*{13mm} Crystallogr.} 8:96 (1974). \\
44. I. Tsatskis (in preparation). \\
45. J. Kulik, D. Gratias, and de D. Fontaine, {\em Phys. Rev. B}, 
40:8607 (1989). \\
46. A. Finel and D. de Fontaine, {\em J. Statist. Phys.} 
43:645 (1986). \\
47. D. Le Bolloc'h et al. (private communication and in preparation).